\documentclass[11pt]{article}%
\usepackage{hyperref}
\usepackage{cite}
\usepackage{amsmath}
\usepackage{amsfonts}
\usepackage{amssymb}
\usepackage{graphicx}
\usepackage{amscd}%
\ifx\pdfoutput\relax\let\pdfoutput=\undefined\fi
\newcount\msipdfoutput
\ifx\pdfoutput\undefined\else
\ifcase\pdfoutput\else
\ifx\paperwidth\undefined\else
\ifdim\paperheight=0pt\relax\else\pdfpageheight\paperheight\fi
\ifdim\paperwidth=0pt\relax\else\pdfpagewidth\paperwidth\fi
\fi\fi\fi
\begin{document}

\title{Dark matter and dark energy from pockets of gravity created by quantum
tunneling of the inflaton potential}
\author{M. Chaves\thanks{Email: mchaves@fisica.ucr.ac.cr}\\\textit{Escuela de Fisica}\\\textit{Universidad de Costa Rica}\\\textit{San Jos\'{e}, Costa Rica}}
\date{December 12, 2013}
\maketitle

\begin{abstract}
It is usually assumed that during the reheating period of inflation the
inflaton field behaves coherently and follows a classical path. It is pointed
out that a background inflaton field $\phi_{0}$ falling \emph{down} a
potential density $V(\phi)$ can and frequently should undergo a quantum
tunneling effect involving an energy $\Delta V\ll V,$ so that it becomes
possible to create perturbative inflatons $\phi^{\prime},$ which, in turn, can
disintegrate into other particles. We argue for this scenario using large
field inflation assumed during a typical slow-roll regime, and typical values
for the physical quantities associated with the tunneling effect are worked
out. It is shown how, as a result of these highly energetic and localized
processes, the components of the metric receive strong gradients within a
small volume. These pockets of gravity can gravitate both attractively or
repulsively, since the Ricci tensor contains squared partial derivative terms
with either positive or negative sign, so they can be a component of either
dark matter or dark energy. As the Universe expands the curvature of the
pockets weakens as $1/a^{2},$ where $a(t)$ is the scale factor of the
expansion, but it is shown that, as long as the integral over the gravitating
gravitational fields of the pocket is done over all of the expanding volume of
the pocket, the value of the integral is not affected by the expansion.
Geometrically speaking, the gravitation of a pocket is produced by the
convolutions of the metric inside it; as the Universe expands the convolutions
are over a much large volume so they are flatter, but an integration over the
larger volume gives the same result as before the expansion. It is just a
change of scale of the same geometry. A common cause (inflaton quantum
tunneling) for both dark matter and dark energy explains why their quantities
are of the same order of magnitude.


\end{abstract}

\section{Introduction.}

Although initially it was not clear whether dark matter was a real substance
or a modification of celestial mechanics, in the last few years it has been
established as a real substance, thanks to the technique of gravitational
lensing.\cite{DM} Even cosmological-size structures made of dark matter have
also been studied using this technique.\cite{scaffolding} Similar doubts occur
with respect to another peculiar substance, dark energy, but here the
situation is more uncertain and there are several possible explanations.

In this paper I point out that the general theory of relativity theoretically
contains both dark matter and dark energy. The Ricci tensor is a sum of
several thousand terms, each one the product of several metric components
multiplied by either: 1) one more metric component and a second-order partial
derivative of another component; or, 2) two factors, each one a first-order
partial derivative of a metric component. All terms are divided by the same
common denominator, a polynomial composed of metric components. Some of these
terms can be taken to the right-hand side of Einstein field equations and be
understood as gravitating. There are gravitating terms that are positive and
others negative, so gravity can be a source of dark matter and dark
energy.\cite{MC} The question that arises is, what could be the source of such
energetic metric components?

The gravitational field produced by present day matter distributions is far
too weak to itself gravitate with more than infinitesimal strength (excluding
very unusual and uncommon locations), so that at first sight it would seem
that this theoretical possibility would not have a practical relevance. In the
Einstein field equations the stress tensor density has to be multiplied by the
gravitational constant $G,$ which is very small. The resulting gravitational
field is numerically very small, and its square would be even smaller.

Suppose now that there is some kind of primordial event that produces a large
amount of energy within a small volume. This would imply, from Einstein field
equations, that the Ricci tensor is going to have large gradients over the
mentioned volume. The Ricci tensor is nonlinear in the metric components and
will in turn generate gravitational fields. Notice these terms do not have to
be multiplied by $G$ before they gravitate.

The reader will probably wonder about the weakening effect that the
cosmological expansion will have on such localized gradients. A typical
gravitating term is, for instance, $(g_{\mu\nu,i}/a)^{2},$ where $a(t)$ is the
scale factor in the cosmological expansion. Geometrically, this means that the
cosmological expansion flattens out curvature details. However, the
gravitational potential produced by this gradient squared term is the result
of integrating over all the volume taken by the field, and we shall show that
this integral does not depend on the cosmological expansion. As a result the
gravitational field produced by a gradient squared term does not weaken with
the cosmological expansion as long as we integrate over the new expanded volume.

The question arises if there was a mechanism that could have actually produced
this type of primordial objects. According to our present understanding of the
early Universe, during the slow-roll regime the inflaton field $\phi$ falls
down a potential $V(\phi)$ in a space-independent but time-dependent manner,
that is, as a function $\phi=\phi(t).$ It is acting as a coherent aggregate of
inflatons. As the field falls down the potential $V(\phi)$ it follows a
classical equation of motion generated by the effective action of the path
integral. The reheating of the Universe is carried out with the help of
additional fields which also follow classical equations of motion.

In the high energy standard model there is a similar scalar field, the Higgs
boson $\Phi,$ which is thought to generate mass through a constant nonzero
uniform value $\Phi_{0}$ it possesses in the vacuum. (Technically the Higgs,
unlike the inflaton which is real, is a complex $SU(2)$ isospin doublet field,
but this difference is immaterial for our purposes.) In this model there
exists a potential $V(\Phi)$ that is supposed to have a minimum at a
\textit{nonzero} value of the $\Phi_{0}$. The coherent behavior can be best
worked out through the Feynman path integral formalism. The amplitude is the
sum over all possible paths of the field $\Phi(x),$ starting from an initial
configuration $\Phi(t_{i,}\mathbf{x})$ and ending in a final configuration
$\Phi(t_{f},\mathbf{x})$. In general, in the case of a background field, the
classical path is a space-independent solution $\Phi(t)$ generated by the
effective action of the integral.\cite{Pokorski} In the standard model of high
energy physics, the potential $V(\Phi)$ has a minimum for a \textit{constant
and uniform} value $\Phi=\Phi_{0}$. This value is the vacuum expectation value
of the quantum field $\hat{\Phi}$, so that $\left\langle 0|\hat{\Phi
}|0\right\rangle =\Phi_{0}.$ (Here, for the sake of simplicity, we disregard
the doublet nature of the Higgs.) According to quantum field theory, there
should exist a perturbative particle $\Phi^{\prime}$ (also dubbed the Higgs, a
case of abuse of language) which represents the quantum fluctuations of the
Higgs about this minimum at $\Phi_{0}.$ Recent observations at
CERN\cite{H1,H2} strongly suggest that this perturbative Higgs particle
$\Phi^{\prime}$ does indeed exist. This makes a strong case for the objective
reality of our ideas about background fields and nonzero vacuum expectation
values, and, in general, the use of the path integral formalism in these
cases. So we shall use this formalism for case of the inflaton, and follow its
logical consequences.

There is an important difference, very pertinent to us, between the physical
situations of the inflaton and the Higgs scalar fields.\ The Higgs is stably
sitting at a minimum of the potential $V(\Phi),$ while the inflaton is sliding
down an inclined wall of the potential $V(\phi).$ So the state $\Phi=\Phi_{0}$
of the Higgs field corresponds to a vacuum state, where by vacuum state we
mean the quantum state of least energy. On the other hand, the state
$\phi=\phi_{0}(t)$ of the inflaton field is not a vacuum state, at least not
in the sense of being a state of least energy. If the inflaton potential
$V(\phi)$ has a self-coupling term $\lambda\phi^{4}$ it would allow a quantum
amplitude for the disintegration of an inflaton. The existence of
disintegration channels open to the Higgs and the inflaton does not imply that
any disintegration is actually going to occur, because there may no be no
energy available for particle creation. Take, for example, a case in the high
energy standard model: The mass of the neutral vector boson $Z_{0}$ of the
standard model is generated through the coupling $\lambda\Phi_{0}^{2}Z_{0}%
^{2},$ where $\lambda$ is a coupling constant. Thus there is a nonzero quantum
amplitude for the process $\Phi_{0}\rightarrow\Phi Z_{0}Z_{0},$ but this
process is impossible because there is no available source for the required
energy since the Higgs is impeded of taking any value other than $\Phi_{0}$.
Going back to inflaton in the slow-roll regime, assume the system is falling
down the potential $V(\phi)$. Then the inflaton quantum field can disintegrate
by means of the quartic self-coupling that enables the channel $\phi
\rightarrow\phi\phi\phi.$ But now, unlike the previous case with the Higgs,
there is the possibility that the quantum tunneling effect can supply the
energy necessary for the disintegration. This would happen as follows: in a
small volume $\mho$ the value of the inflaton field drops from $\phi_{0}$ to,
say, $\phi_{0}-\Delta\phi_{0},$ so that there is an energy $\mho\Delta V=\mho
V^{\prime}(\phi_{0})\Delta\phi_{0}$ available for the creation of the new
perturbative inflaton fields. Here $V^{\prime}=dV/d\phi.$ The tunneling effect
is likely to happen when the system is sliding \emph{down} the potential. If
the system is going \emph{up} (losing kinetic energy as it does) the path
integral would greatly inhibit the quantum tunneling (as we shall see) and the
effect would be unlikely to happen. The path integral also inhibits the tunnel
effect over large volumes and large changes in the field. The energy that
becomes available this way can be used to create three inflatons, two
perturbative inflatons $\phi^{\prime}$ with a null vacuum expectation value
and another $\tilde{\phi}_{0}$ with a large vacuum expectation value $\phi
_{0}-\Delta\phi_{0}.$ The inflatons product of the disintegration can in turn
disintegrate into other particles and contribute to the reheating.

Quantum tunneling of the potential $V(\phi)$ on a small volume is precisely
what is needed to produce strong localized gravitational fields. Whenever a
tunneling effect occurs, the stress tensor $T_{\mu\nu}$ acquires large
gradients inside the volume $\mho$ which, in turn, produce large gradients in
the Einstein tensor and so, gravity pockets. We proceed to develop these ideas mathematically.

\section{The slow-roll regime.}

In an inflationary model one usually assumes that the inflaton field undergoes
a slow-roll regime during which the Universe expands very fast. At the end of
this period there is the period of reheating, during which the Universe
becomes populated by particles at high temperature. Here we are going to study
certain quantum mechanical aspects pertaining the slow-roll regime and reheating.

Initially the treatment of reheating was done perturbatively assuming a
damping term in the inflaton equation of motion:\cite{DL,AFW,ASTW}%
\begin{equation}
\ddot{\phi}+3H\dot{\phi}+\Gamma\dot{\phi}=-V^{\prime}(\phi). \label{eqmotion}%
\end{equation}
There could be a problem here because the small coupling required by the
slow-roll results in small values for the damping term and therefore very
little particle production during the time scale of the expansion. Under the
assumption that the inflaton is an aggregate of coherent particles which can
be treated classically and as a solely time-dependent field, it was suggested
that the inflaton, appropriately coupled to other helping fields, could
undergo parametric resonance and produce large quantities of secondary
particles.\cite{TB,DK,KLS} Another popular model was tachyonic preheating,
which also results in a rapid process of reheating. In this model the inflaton
is assumed to be a tachyonic field, that is, a field with a negative mass
squared term.\cite{GPR,FGGKLT,FKL} As a result of this sign the long
wavelength modes of the system can undergo exponential growth. There is always
such a tachyonic (spinodal) instability in cases with spontaneous symmetry
breaking. This mechanism allows the production of large massive particles that
can be pertinent to baryogenesis. There are several other reheating models,
most of them hybrid, but we cannot cover them here. All of these models are
classical, in the sense that the fields follow classical equations of motion,
that is, they follow the classical path of the integral.

We shall assume large field\ inflation (also called chaotic inflation), where
the inflaton is a spatially homogenous scalar field $\phi$ which must take
values larger than the reduced Planck energy scale%
\begin{equation}
M_{P}\equiv(8\pi G)^{-1/2},
\end{equation}
as we shall soon see. In the general relativity calculations that follow,
Greek letters $\mu,\nu,\ldots$ take the values $0,1,2,3,$ and Latin letters
$i,j,\ldots$ take the values $1,2,3.$ We assume an effective potential%
\begin{equation}
V(\phi)=\frac{1}{2}m^{2}\phi^{2}+\frac{\lambda}{4!}\phi^{4}, \label{pot}%
\end{equation}
that has a non-tachyonic mass term and a self-coupling $\lambda>0$ which we
take to be sufficiently smaller than unity so as to ensure that the inflaton
quantum field theory can be treated perturbatively.

We take the classical path (the background field in the path integral) to be
$\phi_{0}(t).$ The inflaton can be written as the sum of the classical field
and the perturbative field:%
\begin{equation}
\phi(x)=\phi_{0}(t)+\phi^{\prime}(x). \label{varphi}%
\end{equation}
Here the $\phi^{\prime}(x)$ is the perturbative inflaton quantum field. The
potential $V(\phi)$ and the location (the dot) of the classical path $\phi
_{0}$ are depicted in Fig. 1. The arrow points in the direction of the moving
field. At the origin there is another dot, to indicate the final state of the
system after reheating. This state would be a true vacuum and quantum
tunneling will be impossible once the system is there.%
\begin{figure}
[h]
\begin{center}
\includegraphics[
natheight=2.402800in,
natwidth=3.477100in,
height=2.4028in,
width=3.4771in
]%
{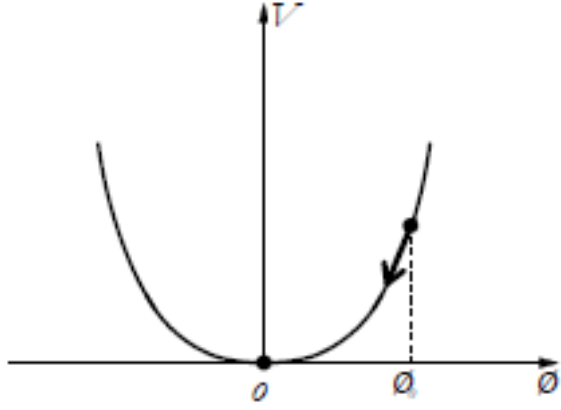}%
\caption{The potential $V(\phi)$. The state of the system is indicated by a
dot, and is falling, as the arrow indicates. The other dot at the origin shows
what would be a true vacuum. }%
\end{center}
\end{figure}

In order for the expansion of the Universe to last several $e$%
-foldings\cite{LL} it is necessary that the following two conditions are met:
\[
\frac{M_{P}^{2}}{2}\left(  \frac{V^{\prime}(\phi)}{V(\phi)}\right)  ^{2}%
\ll1,\quad M_{P}^{2}\left\vert \frac{V^{\prime\prime}(\phi)}{V(\phi
)}\right\vert \ll1.
\]
We have assumed that the inflaton potential can be treated perturbatively, so
for the following calculations the neglect the interactive term $\frac
{\lambda}{4!}\phi^{4}$ with respect to the mass term $\frac{1}{2}m^{2}\phi
^{2}.$ Then the two relations above basically contain the same information and
can be rewritten as one, using the initial value of the field:
\begin{equation}
\epsilon=\frac{2M_{P}^{2}}{\phi_{0}^{2}}\ll1. \label{epsilon}%
\end{equation}

It is usually assumed that 60 $e$-foldings are required to have an adequately
inflated Universe. With this information we can further restrict the value of
the $\epsilon$ parameter:%
\begin{equation}
\frac{a(t_{2})}{a(t_{1})}=\exp\left(  -\int_{\phi_{0\,}}^{0}8\pi G\frac
{V(\phi)}{V^{\prime}(\phi)}d\phi\right)  =\frac{\phi_{0}^{2}}{4M_{P}^{2}%
}=\frac{1}{2\epsilon}\sim60, \label{folds}%
\end{equation}
so that%
\[
\epsilon\sim\frac{1}{120}.
\]
Going back to equation (\ref{epsilon}) with this result we obtain:%
\begin{equation}
\phi_{0}\sim15.5M_{P}. \label{super}%
\end{equation}
In this scenario inflaton fields are superPlanckian, even if the energy
density is not.

To keep within the margins of a classical treatment of gravity it is necessary
that the energy available in a small Planck volume is less than the Planck
energy $M_{P},$ or $V(\phi)M_{P}^{-3}\ll M_{P}.$ We define the parameter $\xi$
as follows:%
\begin{equation}
\xi=V(\phi_{0})/M_{P}^{4}\lesssim\frac{1}{10}, \label{scale}%
\end{equation}
to ensure a basically classical treatment of gravity. It may be that this
value has to be even smaller. From (\ref{scale}), (\ref{super}) and
(\ref{epsilon}), we conclude the following for the inflaton's mass:%
\begin{equation}
\epsilon\xi=\frac{m^{2}}{M_{P}^{2}}\sim\frac{1}{1200}, \label{chain}%
\end{equation}
or $m\sim M_{P}/35.$

We can also easily get the period $\tau$ of an $e$-folding using the Friedmann
equation:%
\begin{equation}
\frac{1}{\tau}\equiv H=\sqrt{\frac{8\pi G}{3}\rho}\sim\frac{M_{P}^{-1}}%
{\sqrt{3}}\sqrt{\xi M_{P}^{4}}\lesssim\frac{M_{P}}{5.5},
\end{equation}
or $\tau\gtrsim5.5M_{P}^{-1}.$ If we use a smaller value for $\xi,$ say 1/100,
then $\tau\sim17M_{P}^{-1}.$

\section{ The path integral: alternative quantum paths.}

The path integral for the inflaton problem (assuming classical gravity and
omitting hybrid fields and the fields that are to be obtained from secondary
disintegrations) is:%
\begin{equation}
I=\int D\phi\exp\left[  i\int_{t_{\,i}}^{t_{f}}dt\int d^{3}x\sqrt
{-g}\mathcal{L}[\phi,g]\right]  \label{integral}%
\end{equation}
where the Lagrangian density $\mathcal{L}$ is given by
\begin{equation}
\mathcal{L}=-\frac{1}{2}\partial_{\mu}\phi g^{\mu\nu}\partial_{\nu}\phi
-V(\phi)+\frac{1}{16\pi G}R.
\end{equation}
We are using the metric signature $(-+++).$ The symbol $\int D\phi$ is telling
us to sum over all paths of the field $\phi(x)$, but there is no similar sum
over all paths of the metric $g=(g_{\mu\nu})$ since we are taking gravity to
be classical. Its time development simply follows the equations of motion of
the theory of general relativity. The path integral goes from from an initial
configuration at time $t_{i}$ of the fields $\phi(t_{i},\mathbf{x})$ and
$g_{\mu\nu}(t_{i},\mathbf{x)}$ to a final one at time $t_{f}$ when $\phi
(t_{f},\mathbf{x})=0$ (and thus also $V=0$ for the potential we have chosen)
and the metric has a value $g_{\mu\nu}(t_{f},\mathbf{x)}$ that follows from an
application of the Einstein field equations.

The integral for the Robertson-Walker spacetime is defined over a countable
number of paths as follows. Construct a lattice in spacetime between the two
hypersurfaces $t=t_{1}$ and $t=t_{2}$ by dividing it into equal volume
elements $\Delta t\Delta^{3}x$ (using space comoving coordinates). Each point
of the lattice is identified by the set of integers $n=(n_{0},n_{1}%
,n_{2},n_{3})=(n_{0},\mathbf{n}).$ The path integral is then defined by the
following limit:
\begin{equation}
I\equiv\lim_{\Delta t\Delta^{3}x\rightarrow0}%
{\displaystyle\prod\limits_{n}}
\int_{-\infty}^{+\infty}d\phi_{n}\exp\left(  i\mathcal{L}[\phi_{n},g_{n}%
]\sqrt{-g_{n}}\Delta t\Delta^{3}x\right)  ,
\end{equation}
where it is understood that there is one integral $\int d\phi_{n}$ for each
vertex of the lattice. We call each of these integrals a \textit{product
integral}. The Lagrangian density for the inflaton field in the path integral
is given by%
\begin{equation}
\mathcal{L}\Delta t\Delta^{3}x=-\frac{1}{2}\frac{\left(  \phi_{n_{0}%
+1,\mathbf{n}}-\phi_{n_{0},\mathbf{n}}\right)  ^{2}}{\Delta t}g^{00}\Delta
^{3}x+\cdots-V(\phi_{n})\Delta t\Delta^{3}x, \label{KE}%
\end{equation}
where the ellipsis indicates the three other similar kinetic energy terms. It
is obvious that every one of the four kinetic energy terms in (\ref{KE}) has a
different one of the intervals $\Delta t,\Delta x,\Delta y,\Delta z$ as denominator.

The difference between the classical and quantum regimes becomes clear from an
inspection of one of the factor integrals above:%
\begin{equation}
\int_{-\infty}^{+\infty}d\phi_{n}\exp\left(  i\mathcal{L}[\phi_{n},g_{n}%
]\sqrt{-g_{n}}\Delta t\Delta^{3}x\right)  . \label{FI}%
\end{equation}
The quantity%
\begin{equation}
\Delta E_{\mathcal{L}}\sim\mathcal{L}[\phi_{n},g_{n}]\sqrt{-g_{n}}\Delta^{3}x
\end{equation}
gives the scale of the energy of the system, and $\Delta t$ the time scale. A
typical classical system has $\Delta E_{\mathcal{L}}\Delta t\gg1,$ so that the
integrand of the imaginary exponential is a large quantity that is going to
produce a rapid sinusoidal oscillation and thus result in a null value for the
integral. The only exception to this behavior is when the Lagrangian does not
vary as the field $\phi_{n}$ varies. This is precisely what happens for the
classical path, where%
\begin{equation}
\delta\mathcal{L=}\frac{\delta\mathcal{L}}{\delta\phi}\delta\phi
=[\partial_{\mu}(g_{\mu\nu}\partial_{\nu}\phi)-V^{\prime}(\phi)]\delta\phi=0,
\end{equation}
the Euler-Lagrange equation.

Notice that the concept of \textit{path} only makes sense because of the
kinetic energy terms in (\ref{KE}), which relates the value of the field
$\phi$ in two neighboring points of the lattice. (If the Lagrangian consisted
only of the potential term $V(\phi)$ each integral (\ref{FI}) would result in
the same value.) But the kinetic energy terms, like, for example,%
\begin{equation}
-\frac{1}{2}\left(  \phi_{n_{0}+1,\mathbf{n}}-\phi_{n_{0},\mathbf{n}}\right)
^{2}g^{00}\Delta^{3}x/\Delta t, \label{deno}%
\end{equation}
relate two different product integrals (in this case, the integrals with
respect to $\phi_{n_{0}+1,\mathbf{n}}$ and $\phi_{n_{0},\mathbf{n}})$.

In the case that $\Delta E_{\mathcal{L}}\Delta t\gtrsim1,$ the values of the
argument of the imaginary exponential are close to unity so that there is not
rapid oscillation and many different paths contribute. This does not mean that
all paths contribute equally, since all the ones that involve a sudden change
of the field in time or space will still get more strongly suppressed. For
example, in the expression (\ref{deno}) for the kinetic energy with respect to
time, it can be seen that the time increment $\Delta t$ is dividing, not
multiplying, and so enhancing the effect of any abrupt change $\left(
\phi_{n_{0}+1,\mathbf{n}}-\phi_{n_{0},\mathbf{n}}\right)  ^{2}.$ As the
interval $\Delta t$ gets smaller, the abrupt change keeps its value (being
abrupt), and the argument of the exponential get large and so the exponential
oscillates very fast.

\section{Inflaton decay and the quantum tunnel effect.}

The Higgs scalar $\Phi$ of the high energy standard model has in the vacuum a
uniform, constant nonzero value $\Phi_{0}.$ It is believed that $\Phi=\Phi
_{0}$ is a local or perhaps an absolute minimum of some potential $V(\Phi)$.
(As we wrote before, the Higgs field is a doublet of $SU_{I}(2),$ but we are
not going to worry about this irrelevant technical aspect.) This uniform,
constant value is the vacuum expected value of the Higgs or $\left\langle
0|\Phi|0\right\rangle =\Phi_{0}$. But the case of the inflaton $\phi$ in the
slow-roll regime is rather different: $\phi=$ $\phi_{0}(t)$ is not a vacuum
state (that is, the lowest energy state), but is instead a uniform,
time-dependent, solution of the equation of motion, that is, a background
field. The function $\phi_{0}(t)$ is sliding down the inclined curve of the
$V(\phi)$ potential, as was seen in Fig. 1. If the system were at the origin
$O$ of Fig. 1, it would become totally stable and the quantum tunneling would
not be possible anymore. As usual there are quantum perturbations
$\phi^{\prime}(x)$ related to the classical value $\phi_{0}(t)$ of the
inflaton by equation (\ref{varphi}).

The self-interactive term $\lambda/4!\phi^{4}$ present in the potential
$V(\phi)$ allows for the possibility that the inflaton field $\phi$
disintegrates into three inflatons. In this case the original background
inflaton $\phi_{0}$ cannot change too much (because the path integral
suppresses abrupt changes, as was pointed out in last section), so that there
will still have to be a background field we shall call $\tilde{\phi}_{0}%
=\phi_{0}-\Delta\phi_{0}$. The process is then%
\begin{equation}
\phi_{0}\rightarrow\tilde{\phi}_{0}\phi^{\prime}\phi^{\prime},
\end{equation}
corresponding to the Feynman diagram $-i(\lambda/4!)(2\pi)^{4}\delta
(p_{1}-p_{2}-p_{3}-p_{4}),$ where the $p_{i},$ $i=1,2,3,4,$ are the 4-momenta
of the inflaton fields. The 3-momenta $\mathbf{p}_{2}\mathbf{+p}%
_{3}+\mathbf{p}_{4}\mathbf{=0}$ can always sum to zero (even if $\mathbf{p}%
_{2}\mathbf{=0)}$ so 3-momentum conservation does not constitute a problem,
but we do need a source of energy to ensure energy conservation. A process
like this one cannot occur when the system is at the minimum of the potential
because there is simply no source for the energy required for the creation of
the perturbative inflatons. Such is the case with the Higgs field, where it is
assumed that the system is at a minimum of the potential $V(\Phi),$ so there
is no value of $\phi$ the system can jump to liberate potential energy. It is
a real vacuum so it is stable. But in the case of the inflaton, a small change
$-\Delta\phi_{0}$ in the inflaton leads to a drop $\Delta V(\phi
_{0})=V^{\prime}(\phi_{0})(-\Delta\phi_{0})$ in the potential energy, and this
energy becomes available for inflaton production.

Consider the change in path that occurs when within a 3-volume $\mho$ with
rounded edges the value of the inflaton falls from $\phi_{0}$ to $\tilde{\phi
}_{0}=\phi_{0}-\Delta\phi_{0}.$ The change in the potential energy of the
inflaton is%
\begin{align}
\Delta E_{\mho}  &  =-\mho\Delta V=-\mho V^{\prime}\Delta\phi_{0}=-2\mho
V\Delta\phi_{0}/\phi_{0}\nonumber\\
&  =-2\mho\xi M_{P}^{4}\Delta\phi_{0}/\phi_{0}, \label{deltaE}%
\end{align}
which then becomes available for the disintegration $\phi\rightarrow\phi
\phi\phi.$ This process can occur as long as it happens in the time $\Delta t$
given by the uncertainty principle%
\begin{equation}
\Delta E_{\mho}\Delta t\sim1/2. \label{u.p.}%
\end{equation}

The path modification in the path integral that is involved is illustrated in
Fig. 2. In this figure we take as independent variables one of the three space
coordinates (the $x)$ and the negative of the inflaton field $-\phi,$ and, as
dependent variable, the potential $V.$ We see the surface of the potential $V$
and how it is inclined in the direction of $-\phi.$ In the $x$-direction the
surface does not change until, for a certain value $-\phi,$ the surface caves
down a depth $\Delta V$ with a width $\Delta x$. As the field keeps falling in
the $-\phi$ direction, a surface like a ditch is formed. If we could draw the
figure with two more dimensions visible, the ditch would occur also in the
$y$- and $z$-direction: it is basically a sphere within which the potential
has diminished by $-\Delta V.$%
\begin{figure}
[h]
\begin{center}
\includegraphics[
natheight=3.002000in,
natwidth=3.756700in,
height=3.002in,
width=3.7567in
]%
{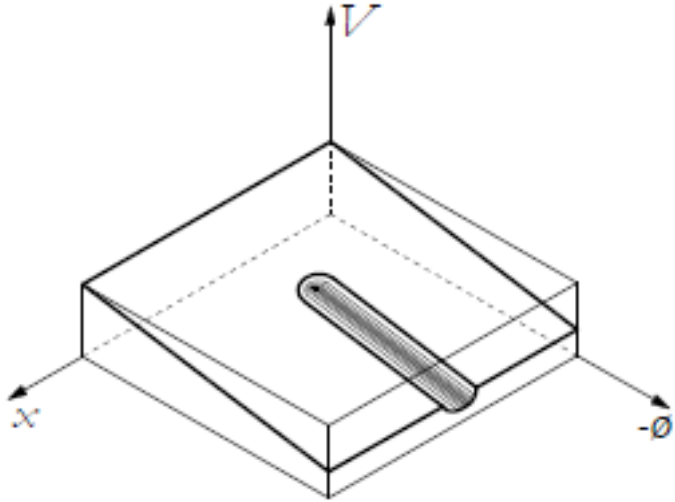}%
\caption{Depiction of the drop $\Delta V$ in potential within a small rounded
volume $\mho$. The value of $\phi$ is diminishing in $\phi$-axis direction and
the width of the volume is shown in the $x$ direction.}%
\end{center}
\end{figure}

This effect bears some resemblance to the Hawking effect which occurs at the
event horizon of a black hole, where a pair is produced with energy
substracted from the gravitational potential. In our case the energy comes
from the inflaton's background field potential energy.

The existence of quantum tunneling has two important consequences. First, it
is a new channel to accelerate the conversion of the inflaton's potential
$V(\phi)$ into radiation, since the perturbative inflatons that are created
within the small rounded volume $\mho$ disintegrate into other particles. For
the reasons detailed in last section the drops in potential $\Delta V$ have to
be small but frequent during the slow-roll. Second, within a small volume
$\mho$ there have been large variations of energy and momentum, a situation
that creates gravity pockets since Einstein field equations relate stress
tensor gradients with gradients in the metric components. After the inflaton
gradients have left their imprint on the metric gradients, the field $\phi$
develops according to its equations of motion, but to us what is important is
what happens with that imprint left on the metric gradients.

\section{Mathematical details of the quantum tunneling.}

For simplicity we are going to assume that the volume $\mho$ is basically a
sphere with diameter $\Delta x,$ so that the relation between these two
quantities is given by $\mho=4\pi(\Delta x/2)^{3}/3\sim\Delta x^{3}/2.$ It is
possible to estimate the time scale $\Delta t$ of the tunneling process from
the time it takes the virtual inflaton quanta to traverse the diameter $\Delta
x.$ Since we know from equation (\ref{chain}) that the mass of the inflaton is
quite small, it can be neglected at the energies involved and take the virtual
inflatons to be massless. We conclude the time scale of the tunneling is given
by $\Delta t\sim\Delta x$. We now insert this time scale into the uncertainty
principle (\ref{u.p.}) and, in the resulting equation, eliminate the tunneling
energy $\Delta E_{\mho}$ by means of (\ref{deltaE}). This way we obtain the
relation $\Delta x^{4}\left(  \xi M_{P}^{4}\right)  \Delta\phi_{0}/\phi
_{0}\sim1/2,$or%
\begin{equation}
\Delta x=M_{P}^{-1}\left(  2\xi\Delta\phi_{0}/\phi_{0}\right)  ^{-1/4}.
\label{deltax}%
\end{equation}

We can use the path integral to obtain a relation between $\Delta x$ and
$\Delta\phi_{0}$, using the fact that the path integral suppresses abrupt
changes in the field due to the kinetic energy term. Consider a classical path
that comes from the past and arrives at the lattice point $n=(n_{0}%
,\mathbf{n);}$ there the variable of integration is $\phi_{n}.$ Of the four
covariant kinetic energy terms, we shall only treat the one that has the
partial derivatives with respect to time, since the other cases are very
similar. To avoid a plethora of subindices for the case chosen we will simply
write $\phi_{n_{0}}$ for each variable of integration, using only the time
lattice coordinate. The next lattice point in time is $n_{0}+1,$ with a
variable of integration $\phi_{n_{0}+1},$ and occurs after a period $\Delta
t.$ We assume that the tunneling process occurs during this period, and, as
discussed above, that $\Delta t\sim\Delta x$ (and, as a matter of fact,
$\Delta x\sim\Delta y\sim\Delta z$)$.$ We assume that there is a small jump
$-\Delta\phi$ in the value of the field during the time $\Delta t,$ and call
the new value $\phi_{n_{0}+1}.$ We also assume after this time the development
of the field again follows a classical path. Thus the quantum tunneling is a
jump in the value of the field between two segments of classical paths. (See
Fig. 3.)%
\begin{figure}
[h]
\begin{center}
\includegraphics[
natheight=2.639900in,
natwidth=2.863400in,
height=2.6399in,
width=2.8634in
]%
{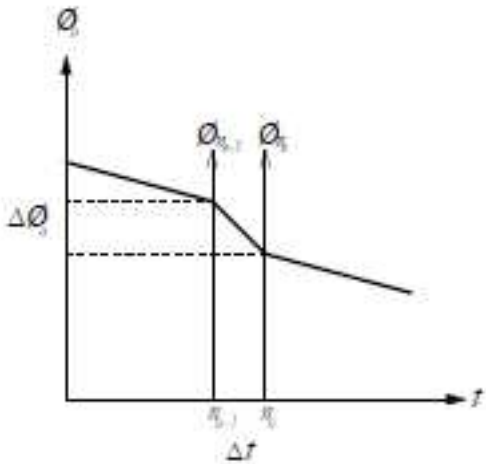}%
\caption{A classical path comes from the left, then there is quantum tunneling
between times $n_{0}-1$ and $n_{0}$, then another classical path proceeds on
to the right.}%
\end{center}
\end{figure}
Then the product integral is%
\begin{equation}
\int_{-\infty}^{+\infty}d\phi_{n_{0}+1}\exp\left(  -i\frac{1}{2}\left(
\phi_{n_{0}+1}-\phi_{n_{0}}\right)  ^{2}\Delta x^{2}\right)  .
\end{equation}

The question one has to ask here is: Just how fast is the field $\phi
_{n_{0}+1}$ changing in a neighborhood of the new value of $\phi_{0}$ after
the drop in value $\Delta\phi_{0}$ of the inflaton? To find that out we take
the functional derivative:%
\begin{equation}
-\frac{1}{2}\frac{\delta\left[  \left(  \phi_{n_{0}+1}-\phi_{n_{0}}\right)
\Delta x^{2}\right]  }{\delta\phi_{n_{0}+1}}\delta\phi_{n_{0}+1}=-\phi
_{0}(t)\Delta\phi_{0}\Delta x^{2}\frac{\delta\phi_{n_{0}+1}}{\phi_{0}(t)},
\end{equation}
and conclude that how fast the exponential is oscillating depends on the
coefficient $\phi_{0}(t)\Delta\phi_{0}\Delta x^{2}.$ If $\Delta\phi_{0}$ is
very small, we basically have the classical solution, with no break in the
path, and a rather small amount of energy available for perturbative inflaton
production. If we allow the imaginary exponential to have 10 oscillations per
unit that will result in a small amplitude, just how small would depend on the
regularization procedure to ensure convergence of the factor integral. Our
goal is to obtain an estimate of the energies and sizes involved in the
tunneling process. We assume%
\begin{equation}
\phi_{0}(t)\Delta\phi_{0}\Delta x^{2}=10, \label{value}%
\end{equation}
and use it in (\ref{deltax}), along with (\ref{super}), to get a value for the
size:%
\begin{equation}
\Delta x=11M_{P}^{-1}. \label{size}%
\end{equation}
The value for $\Delta x$ is rather insensitive to the number chosen in
(\ref{value}). Thus, if we had taken used 20 instead of 10, then $\Delta
x=8M_{P}^{-1}.$

With this value for the size we can use again (\ref{u.p.}), this time to get
an estimate of the energy available in volume $\mho$:%
\[
\Delta E_{\mho}\sim\frac{1}{22}M_{P}.
\]
There is not enough energy density for us to require using quantum gravity in
our calculation. A typical value for the change in the background inflaton due
to the tunnel effect can be obtained immediately from (\ref{value}),
(\ref{super}) and (\ref{size}), with the result%
\begin{equation}
\Delta\phi_{0}=\frac{1}{180}M_{P}. \label{change}%
\end{equation}

It is an interesting question to see if it is possible for a black hole to be
created under these conditions. The Schwarzschild radius of a black hole is
$r_{S}=2Gm,$ where $m$ is the energy of the black hole. In this case
$m=\frac{1}{20}M_{P},$ so that $r_{S}=2Gm=\frac{1}{10}M_{P}^{-1}.$ Thus for a
black hole to be created with the energy available from the tunnel effect its
diameter should be of $M_{P}^{-1}/5,$ while the diameter of the rounded volume
created by the tunnel effect is 55 times that, so it should not produce a
black hole. But our calculations are for a typical tunneling effect. In the
more extreme and unlikely cases there could occur black hole production.

After all the energy density of the inflaton $V(\phi_{0})$ is spent the
potential and the field are both zero. A constant $\phi_{0}=0$ value for the
field is a solution of the classical equations of motion$.$ Thus the system
has gone between two classical solutions. In between, the quantum tunnel
effect accelerates the energy depletion by taking chunks of energy $\mho\Delta
V$ from the inflaton energy density. All this energy is used to produce
perturbative inflaton quanta, which in turn decay into other particles.

\section{The components of the metric and almost-Planckian stress tensors.}

The gravitational field, a quantity which has no units in natural units, has
in general very small values in our Universe. The field due to the sun in the
neighborhood of the earth is of the order of $10^{-8},$ and that due to the
our galaxy near its rim of the order of $10^{-6}.$ So, if one assumes that the
gravitational field created by the gravitational field goes as its square, the
numbers for the induced gravitational field come out to be infinitesimal.

Our Universe is mostly flat, and the metric components have values $g_{\mu\nu
}=\eta_{\mu\nu}+\varepsilon_{\mu\nu},$ where $\eta_{\mu\nu}$ is the Minkowski
metric $\operatorname*{diag}(-1,+1,+1,+1)$ and the $\varepsilon_{\mu\nu}(x)$
are gravitational potentials of the order of $10^{-6}$ or less. This smallness
of the potentials is due to the small and smooth stress tensor densities
available in our Universe today. (We are not taking into consideration
irrelevant special cases like a black hole.) But what has happened in the past
when the stress tensor took very large values?

An alternative form of the Einstein equations is%
\begin{equation}
R_{\mu\nu}=M_{P}^{-2}\left(  T_{\mu\nu}-\frac{1}{2}g_{\mu\nu}T_{\lambda
}^{\lambda}\right)  \equiv M_{P}^{-2}\tilde{T}_{\mu\nu}. \label{Einstein}%
\end{equation}
The Ricci tensor, on the left side of the equation, can be expanded by direct
calculation and results in a sum of 9990 terms, all of which are the product
of eight metrics. This sum is divided by a common denominator which is the
square of the determinant of the metric, that is, the whole sum is divided by
the same polynomial of order eight in the metric components. Furthermore,
every one of the terms in the sum either has one of the metric factors with a
second order partial derivative $g_{\mu\nu,\rho\sigma}$ or, two factors with a
first-order partial derivative each: $g_{\mu\nu,\rho}g_{\alpha\beta,\sigma}$.
That is, Einstein equations can be written in the form%

\begin{equation}
\frac{1}{%
{\displaystyle\sum}
g_{\ast\ast}^{8}}\left(
{\displaystyle\sum}
g_{\ast\ast}^{6}\frac{\partial g_{\ast\ast}}{\partial x^{\mu}}\frac{\partial
g_{\ast\ast}}{\partial x^{\nu}}+%
{\displaystyle\sum}
g_{\ast\ast}^{7}\frac{\partial^{2}g_{\ast\ast}}{\partial x^{\rho}\partial
x^{\sigma}}\right)  =M_{P}^{-2}\tilde{T}_{\ast\ast}, \label{first}%
\end{equation}
where the $g_{\ast\ast}$ are metric components with two subindices and the
tensor $\tilde{T}_{\mu\nu}$ was defined in (\ref{Einstein}).

Naturally there are other ways of writing the terms that make up $R_{\mu\nu}.$
One such way is to take the previous expression and group terms so that the
determinant of the metric appears as a numerator. This factor will then cancel
one of the two determinants in the denominator. Using this type of grouping
there will be terms that have a determinant squared in the denominator and a
product of eight metrics in the numerator, as before, and other simpler terms
that have only one determinant in the denominator and a product of only four
metrics in the numerator. In this presentation there are only 9426 terms. We
will not use in this paper this alternative way of writing $R_{\mu\nu}.$

Suppose there is a system with an energetic stress tensor $\tilde{T}_{\mu\nu
},$ but not quite large enough to produce a black hole (and thus a
singularity) within the time lapse during which we are going to be studying
the system. Let the stress tensor components be of the order $\tilde{T}%
_{\mu\nu}\sim\chi M_{P}^{4},$ where $\chi(x)$ is a pure number in the range
$1>\chi\geq0$. We now take Einstein equations (\ref{first}) and there
substitute the stress tensor we are considering and the change of variables
$y^{\mu}=M_{P}x^{\mu}$. We obtain a form of Einstein equations with no units:%
\[%
{\displaystyle\sum}
g_{\ast\ast}^{6}\frac{\partial g_{\ast\ast}}{\partial y^{\mu}}\frac{\partial
g_{\ast\ast}}{\partial y^{\nu}}+%
{\displaystyle\sum}
g_{\ast\ast}^{7}\frac{\partial^{2}g_{\ast\ast}}{\partial y^{\rho}\partial
y^{\sigma}}\sim\chi%
{\displaystyle\sum}
g_{\ast\ast}^{8}.
\]
Notice the equation is homogenous in the metric components. If we take
$\chi=0$ then all the partials on the left-hand side of the equation are zero,
and a solution for the metric is $g_{\mu\nu}=\eta_{\mu\nu}$ and we have a flat
spacetime. If we take a typical value for $\chi$ within the galaxy, say
$\chi\sim10^{-6},$ then the diagonal elements of the metric will have the form
$g_{\mu\mu}=\mp1-\varepsilon(x),$ where $|\varepsilon|\sim\chi.$ The
non-diagonal elements are basically zero. If we were to take the limit
$\chi\rightarrow0,$ then $g_{\mu\nu}\rightarrow\eta_{\mu\nu}.$ From a physical
point of view it seems reasonable to assume that for larger values of $\chi,$
say $10^{-3},$ $10^{-2}$ and perhaps even $10^{-1},$ the same scheme
$g_{\mu\mu}=\pm1+\varepsilon(x),$ $|\varepsilon|\sim\chi,$ should still hold
since we still have a basically flat space with a perturbation. In the quantum
tunneling case we are working on, the stress tensor $\tilde{T}_{\mu\nu}$
certainly will have the most complicated forms due to the tunneling so that
the non-diagonal elements of the metric should not be zero. (For example, the
pocket could result with angular momentum in one direction, if the
perturbative inflatons carry out opposite angular momentum so it is
conserved.) So the non-diagonal elements $g_{\mu\nu},$ $\mu\neq\nu,$ should
have values $g_{\mu\nu}\sim\chi,$ at least for some time.

If we keep increasing $\chi$ the equation itself becomes suspect because
quantum gravity and black hole production and radiation have to be taken into consideration.

Let us go back to the case of the quantum tunneling we have been studying. We
want to estimate, for the volume inside $\mho,$ the metric gradients of the
left-hand side of the Einstein equations using the gradients of the tensor
$\tilde{T}_{\mu\nu}$ of the right-hand side. Of course, these gradients are
not the kinetic energies available during the classical segments of the path
(which are negligible), but the transitory kinetic energies produced by the
quantum tunneling, which are%
\[
\tilde{T}\sim\frac{1}{2}\frac{\partial\phi}{\partial x^{\mu}}\frac
{\partial\phi}{\partial x^{\nu}}\sim\left(  \frac{\Delta\phi_{0}}{\Delta
x}\right)  ^{2},
\]
and which occur only within the volume $\mho.$ These gradients can be obtained
from (\ref{change}) and (\ref{size}) above, and, with the help of Einstein
equations, lead to the result:%
\[
\left(  \frac{\Delta g_{\ast\ast}}{\Delta x}\right)  ^{2}=\frac{1}{2000}%
M_{P}^{2}.
\]

These strong gradients at that early time could gravitate strongly. The
question that comes up is if the expansion of the Universe is going to weaken
them to the point of making them ineffective today.

\section{Invariance of the gravitational attraction of gravity pockets, and an
intuitive explanation by means of a toy model.}

As was discussed above, the gravitational fields today are due to the masses
of stars, galaxies and clusters, and are very weak, of the order of $10^{-6}$
or less. The gravitational fields produced by the quadratic terms of these
material sources are negligible. The hypothetical gravitational fields
considered here are due to small primordial pockets of gravity formed by the
imprint of quantum tunneling on the gravitational field. Since today's
gravitational fields, produced by matter, are so weak, it could seem by
analogy that the fields from primordial objects would be very weak, too. Or
perhaps, that they were strong originally, but the expansion of the Universe
rendered them too weak to gravitate in any effective way at the present time.
However, neither of these preconceptions is correct.

It is more convenient now to use a metric which explicitly shows the
Robertson-Walker scale factor. The Robertson-Walker metric is
$\operatorname*{diag}(-1,a^{2},a^{2},a^{2}),$ where $a=a(t).$ We also assume a
gravitational disturbance $g_{\mu\nu}$ in the Robertson-Walker spacetime, so
that the resulting metric is $\bar{g}_{\mu\nu}$:%
\begin{equation}
(\bar{g}_{\mu\nu})=%
\begin{pmatrix}
g_{00} & ag_{01} & ag_{02} & ag_{03}\\
ag_{10} & a^{2}g_{11} & a^{2}g_{12} & a^{2}g_{13}\\
ag_{20} & a^{2}g_{21} & a^{2}g_{22} & a^{2}g_{23}\\
ag_{30} & a^{2}g_{31} & a^{2}g_{32} & a^{2}g_{33}%
\end{pmatrix}
. \label{mixed}%
\end{equation}
We denote by $x=(t,\mathbf{x)},$ $\mathbf{x=(}x^{1},x^{2},x^{3}),$ the
comoving coordinates, and by $y=(t,\mathbf{y),}$\textbf{ }$\mathbf{y=}%
(y^{1},y^{2},y^{3})=a\mathbf{x,}$ the physical coordinates. The gravity pocket
would be a local disturbance in the metric $g_{\mu\nu}=g_{\mu\nu}(x)=g_{\mu
\nu}(t,\mathbf{x}).$

We are going to analyze the expansion of the gravity pockets. The original
imprint due to quantum tunneling is very energetic and varied. The internal
dynamics of the pockets will in some instances make them expand, or keep their
size or shrink. Here we will make the assumption that they expand along with
the cosmos. We are not that interested on the exact evolution of the pockets,
but on whether or not their gravitational effect weakens when as they expand.

We split the expansion period of the pocket from its inception to the present
in two:\ 1) an initial exponential inflationary period when $a\sim\exp
(t/\tau);$ and, 2) an ensuing matter-dominated period with a power law
$a=(tH_{0})^{2/3}$. Here $H_{0}$ is today's Hubble parameter, and we shall use
the fact that $H_{0}^{-1}\gg\tau$. (In other words, the expansion during the
matter-dominated period is too slow compared with local dynamics of the system.)

\textit{Period 1.} Let us denote by $t_{1}$ the time when the pocket was
created (which is also about the same time it left the horizon since $\Delta
t\sim\tau)$, and by $t_{2}$ the time when it reentered it. Also assume that
$a(t_{1})=1.$ After the pocket leaves the horizon shortly after its creation,
causality will only allow changes in the pocket that are restricted to very
small volumes, as compared with the total volume of the pocket. For this first
period we shall simply assume that the values of the metric components have
not changed, and that all that has happened to them is that they have
expanded. Thus if the metric at time $t=t_{1}$ was $\bar{g}_{\mu\nu}%
(t_{1},\mathbf{x)}=g_{\mu\nu}(t_{1},\mathbf{x),}$ then at time $t=t_{2}$ its
components are going to be $\bar{g}_{00}(t_{2},\mathbf{x)}=g_{00}%
(t_{1},\mathbf{x),}$ $\bar{g}_{0i}(t_{2},\mathbf{x)}=a(t_{2})g_{0i}%
(t_{1},\mathbf{x)}$ and $\bar{g}_{ij}(t_{2},\mathbf{x)}=a^{2}(t_{2}%
)g_{ij}(t_{1},\mathbf{x).}$

Explicit calculation with a computer of the $R_{00}$ tensor component for
metric $\bar{g}_{\mu\nu}$ gives a sum of about 12000 terms. As before, each
term has a numerator, which is the product of eight metric components
$g_{\mu\nu}$, and a denominator, which is a polynomial sum of many terms each
one the product of eight metric components $g_{\mu\nu}$. But there are added
complications because now there can be powers of the $a$ in the denominators:
$1$, $a$ and $a^{2}$; and there can be other factors in the numerator: $1,$
$\dot{a},$ $\dot{a}^{2},$ and $\ddot{a}.$ By far most of the terms are of one
of the next six types:%
\begin{align}
&  g_{\ast\ast}^{6}\frac{\partial g_{\ast\ast}}{\partial t}\frac{\partial
g_{\ast\ast}}{\partial t},\,g_{\ast\ast}^{6}\frac{\partial g_{\ast\ast}%
}{\partial t}\frac{\partial g_{\ast\ast}}{\partial ax^{i}},\,g_{\ast\ast}%
^{6}\frac{\partial g_{\ast\ast}}{\partial ax^{i}}\frac{\partial g_{\ast\ast}%
}{\partial ax^{j}},\,\nonumber\\
&  g_{\ast\ast}^{7}\frac{\partial^{2}g_{\ast\ast}}{\partial t^{2}}%
,\,g_{\ast\ast}^{7}\frac{\partial^{2}g_{\ast\ast}}{\partial t\partial ax^{i}%
},\,g_{\ast\ast}^{7}\frac{\partial^{2}g_{\ast\ast}}{\partial ax^{i}\partial
ax^{j}}, \label{types_1}%
\end{align}
all of them divided by the common denominator $D=%
{\displaystyle\sum}
g_{\ast\ast}^{8}=\det^{2}g_{\mu\nu}.$ These terms describe the internal
dynamics of the pockets. There are also other types of terms that include time
derivatives of the scale factor:%
\[
g_{\ast\ast}^{7}\frac{\dot{a}}{a}\frac{\partial g_{\ast\ast}}{\partial
t},\,g_{\ast\ast}^{7}\frac{\dot{a}}{a}\frac{\partial g_{\ast\ast}}{\partial
ax^{i}},\,g_{\ast\ast}^{8}\left(  \frac{\dot{a}}{a}\right)  ^{2},\,g_{\ast
\ast}^{8}\frac{\ddot{a}}{a},
\]
all of them divided by the denominator $D,$ too. These type of terms describe
the relation between the pocket and the expansion of the Universe. If we
consider the Universe as a whole, with gravity pockets contributing to an
average density and pressure, its metric simply is the Robertson-Walker, and
in that case the terms of the type $g_{\ast\ast}^{8}\ddot{a}/a$ sum to give
$-3\ddot{a}/a,$ which is the left-hand side of the Friedmann equation.

The next example, consisting of six terms from $R_{00}$, illustrates why there
is no weakening of the gravitational effect of a gravity pocket during the
first period:%
\begin{equation}
-\frac{g_{\ast\ast}^{7}}{2}\frac{\partial}{\partial(a\mathbf{x})}\cdot
\frac{\partial}{\partial(a\mathbf{x})}g_{00}-\frac{g_{\ast\ast}^{6}}{4}\left(
\frac{\partial g_{00}}{\partial ax}\right)  ^{2}+\cdots=0. \label{sum}%
\end{equation}
Take the volume of the pocket to be $\mho(t_{2})$ at time $t_{2}$, when the
pocket reenters the horizon. Now imagine a sphere $S$ that encloses a volume
$V\gg\mho(t_{2})$, and place the pocket at the center of this sphere. We want
to calculate, from the first term of (\ref{sum}), the gravitational force
field due to the second term. Then:%
\begin{equation}
-\int_{V}\frac{g_{\ast\ast}^{7}}{2}\frac{\partial^{2}g_{00}(t_{1}%
,a(t_{2})\mathbf{x)}}{\partial\lbrack a(t_{2})\mathbf{x]}^{2}}a^{3}%
(t_{2})d^{3}x=\int_{V}\frac{g_{\ast\ast}^{6}}{4}\left(  \frac{\partial
g_{00}(t_{1},a(t_{2})\mathbf{x)}}{\partial a(t_{2})\mathbf{x}}\right)
^{2}a^{3}(t_{2})d^{3}x.
\end{equation}
The gravitational field is related to the metric by $g_{00}=-1-2\mathcal{G},$
and the gravitational force field is%
\[
\mathbf{F}=-\frac{\partial}{\partial\mathbf{y}}\mathcal{G}.
\]
Then, with the help of the divergence theorem $\oint_{\partial V}%
\mathbf{F}\cdot\mathbf{dS=}\int_{V}\nabla\cdot\mathbf{F}dV,$ and since
$V\supset\mho(t_{2}),$ we can write:%
\begin{equation}
-%
{\displaystyle\oint_{S}}
\mathbf{F}\cdot\mathbf{dS}=\int_{\mho(t_{2})}\frac{g_{\ast\ast}^{6}}{4}\left(
\frac{\partial g_{00}(t_{1},a(t_{2})\mathbf{x)}}{\partial a(t_{2})\mathbf{x}%
}\right)  ^{2}a^{3}(t_{2})d^{3}x, \label{P0}%
\end{equation}
from which we conclude that%
\begin{equation}
F=-\frac{1}{4\pi r^{2}}\int_{\mho(t_{2})}g_{\ast\ast}^{6}\frac{1}{4}\left(
\frac{\partial g_{00}(t_{1},x\mathbf{)}}{\partial\mathbf{x}}\right)  ^{2}%
d^{3}x. \label{P1}%
\end{equation}
To go from (\ref{P0}) to (\ref{P1}), we make the change of variable
$\mathbf{y=}a(t_{2})\mathbf{x}$ in the volume integral, then rename
$\mathbf{x}$ the dummy variable $\mathbf{y.}$ Recall now that $a(t_{1})=1,$
and notice the result is exactly the same as it was at time $t_{1}.$ The
gravitational effect of the pocket does not weaken as it expands, as long as
we integrate over the volume of the whole pocket. The toy model at the end of
this section gives another perspective on this result.

\textit{Period 2.} During this period the Universe is expanding under the
power law $a=(tH_{0})^{2/3},$ which corresponds to the dust cosmological
solution. Since the time scale of the cosmological expansion in this case is
so much larger than the time scale of the pocket dynamics, we can basically
neglect in the Einstein equations the terms involving time derivatives of the
scale factor$.$ We proceed as in Period 1, but now we assume that the pocket
is developing according to its internal dynamics. Using the same terms again:%
\begin{equation}
-\frac{g_{\ast\ast}^{7}}{2}\frac{\partial^{2}g_{00}(t,a(t)\mathbf{x)}%
}{\partial\lbrack a(t)\mathbf{x]}^{2}}-\frac{g_{\ast\ast}^{6}}{4}\left(
\frac{\partial g_{00}(t,a(t)\mathbf{x)}}{\partial a(t)\mathbf{x}}\right)
^{2}+\cdots,
\end{equation}
at our present time $t.$ We proceed as before and obtain%
\begin{equation}
F=-\frac{1}{4\pi r^{2}}\int_{\mho(t)}g_{\ast\ast}^{6}\frac{1}{4}\left(
\frac{\partial g_{00}(t,\mathbf{y)}}{\partial\mathbf{y}}\right)  ^{2}d^{3}y.
\label{P2}%
\end{equation}
The main difference between the previous result (\ref{P1}) and this one is
that now the pocket has evolved in time following its own dynamics. The
internal dynamics may change the size of the pocket and the distribution of
the energy among the different metric components but, as long as the integral
is over all its volume, it shows no weakening of its gravitational field due
to the expansion of the Universe.

We can further clarify what is happening by means of a toy model. Consider the
following equation for just one function $g(t,x\mathbf{)}$ with only two
independent variables $(t,x)$ and a scale factor $a(t)$:%
\begin{equation}
\frac{g}{a^{2}}\frac{\partial^{2}g}{\partial x^{2}}+\frac{1}{a^{2}}\left(
\frac{\partial g}{\partial x}\right)  ^{2}=g\frac{\partial^{2}g}{\partial
t^{2}}. \label{third}%
\end{equation}
Notice that every term is of the same order in $g,$ that is, the equation is
homogenous, as is the case in the actual Einstein equations, and that the term
on the right-hand side, the one with partials with respect to time, does not
have scale factors in the denominator, again, as is the case in Einstein
equations. The solution is%
\[
g=\exp\left[  \sqrt{2}t+a(t)x\right]  .
\]
(We are assuming the time scale for cosmological expansion is very large so
that terms with a factor $\dot{a}$ can be neglected when compared with
$\dot{g}$.) Notice how the argument of the solution expands over the
increasing distance it must cover. In a sense, the expansion makes no
difference to the system, except it changes the space scale. It makes sense
the overall gravitation produced by the system would be the same.

We have shown how the gravitational effect of gravity pockets does not weaken
with the expansion of the Universe. Nevertheless, gravity pockets should
weaken, but because of a different reason. For a nearly flat spacetime it is
possible to use the expansion $g_{\mu\mu}=\eta_{\mu\mu}+\varepsilon_{\mu\mu},$
with $\varepsilon_{\mu\mu}\ll1,$ and obtain gravitational waves. The point is
that the spacetime dependent part is very small with respect to one. For a
gravity pocket, a primordial object, the part of the metric that is spacetime
dependent is not so small, so this linearization process is not very accurate.
However, an energetic object such as a gravity pocket will surely emit some
kind of radiation, possibly the quanta of gravity, whatever it may be. This
effect is likely to weaken gravity pockets, but many of them will still be a
source of gravity worth considering.

\section{Gravity pockets acting as dark matter and dark energy.}

It is of interest to have an idea of the size of a gravity pocket at the
present time, disregarding its internal dynamics, that is, strictly on the
basis of the cosmological expansion. We can get a rough estimate as follows.
It has been argued that a gravitational wave created at the Planck time with a
Planck energy $T=2.4\times10^{18}$GeV would nowadays have a frequency\cite{Ma}%
\begin{equation}
f_{0}\sim1.65\times10^{-7}\frac{T}{\text{GeV}}\left(  \frac{g}{100}\right)
^{1/6}=5.8\times10^{11}\text{Hz.}%
\end{equation}
In the above formula we have taken the number of degrees of freedom to be
$g=1000,$ since we are dealing with the reheating period. Due to the sixth
order root the result does no depend much on this choice. Thus the
gravitational wave's wavelength has been red-shifted from $1.6\times10^{-35}$m
to $5.2\times10^{-4}$m, a factor of $3\times10^{31}$ times. The diameter of a
gravity pocket is a few Planck lengths, and it must have increased by a
similar ratio due to the expansion of the Universe. We conclude that the size
of a typical gravity pocket at present time is of a few millimeters.

It has been mentioned above that the numerator of each of the 9990 terms of
$R_{\mu\nu}$ is of one of two types: either it is the product of six metrics
times $g_{\mu\nu,\rho}g_{\alpha\beta,\sigma},$ or of seven metrics times
$g_{\mu\nu,\rho\sigma}.$ The terms with the $g_{\mu\nu,\rho}g_{\alpha
\beta,\sigma}$ product sometimes are squares, sometimes not. Each pocket will
have, at a particular time, a preponderance of some of the metric components.
The ones that are not squares will have positive and negative values and
cancel in the average, so we shall concentrate only on the terms that have
squared partials: $(g_{\mu\nu,\rho})^{2}$.

In the $R_{00}$ there are many terms that are square in the partials, and of
these some are positive and some negative. After the imprint, the gravity
pocket does not have a stress tensor source on the right-hand side of the
Einstein equation, so it is possible to eliminate the common denominator.
Since each term is the product of many metric components and since many of
those factors can be positive or negative, we are going to make some rough
approximations in order to get an idea of the sign of the gravitational
contributions of the pockets. From previous considerations, it is likely that
the spacetime dependent part of the diagonal metric components should be about
1/10 of the constant part, which is $\mp1.$ The off-diagonal terms should be
be also in that range, about 1/10. What we do is to substitute $g_{\mu\nu
}\rightarrow\eta_{\mu\nu}$ in $R_{00},$ except in the factors with partial
derivatives where we leave the differentiated $g_{\mu\nu}$'s intact. This is
not a good approximation but our interest here is to have an idea of how many
of the terms are positive or negative.

After the substitution for the Minkowski metric we are left in the Ricci
tensor $R_{00}$ with 39 are terms with mixed first-order partial derivatives
(that is, they are not squares, and we said that we will ignore them), and 15
first-order partial derivatives that are squares. The latter are:%
\begin{align*}
&  R_{00}\rightarrow\frac{1}{4}g_{11,0}^{2}+\frac{1}{2}g_{12,0}^{2}+\frac
{1}{2}g_{13,0}^{2}+\frac{1}{4}g_{22,0}^{2}+\frac{1}{2}g_{23,0}^{2}+\frac{1}%
{4}g_{33,0}^{2}+\frac{1}{2}g_{10,3}^{2}\\
&  +\frac{1}{2}g_{20,3}^{2}+\frac{1}{2}g_{10,2}^{2}+\frac{1}{2}g_{30,2}%
^{2}+\frac{1}{2}g_{20,1}^{2}+\frac{1}{2}g_{30,1}^{2}-\frac{1}{4}g_{00,1}%
^{2}-\frac{1}{4}g_{00,2}^{2}-\frac{1}{4}g_{00,3}^{2}.
\end{align*}
As we shall see, if the terms with positive sign have preponderance within a
pocket, then it will be a source of dark energy, and if it is the terms with
negative sign that have it, then that it will be a source of dark matter.
There are more terms that are sources of dark energy than of dark matter.

One of the resulting terms is $-\frac{1}{4}g_{00,1}^{2},$ which has a negative
sign, so when we take it to right-hand side of Einstein equation it will act
as a source of gravity:%
\begin{equation}
F=-\frac{1}{4\pi r^{2}}\int_{V}\frac{1}{4}\left(  \frac{\partial
g_{00}(t,\mathbf{x)}}{\partial x^{1}}\right)  ^{2}d^{3}x.
\end{equation}
The force is inwards, and thus, attractive. But a term like $+\frac{1}%
{4}(g_{11,0})^{2},$ that also results from the above-mentioned technique, has
a positive sign, which means that when we take it to the right-hand side of
the Einstein equation it will act as source of \textit{negative }gravity:%
\begin{equation}
F=\frac{1}{4\pi r^{2}}\int_{V}\frac{1}{4}\left(  \frac{\partial g_{11}%
(t,\mathbf{x)}}{\partial t}\right)  ^{2}d^{3}x.
\end{equation}
The field lines this time are going radially towards the pocket, in the
opposite direction as before. A small test mass will experience, as before, a
force against the direction of the field lines, so that in this case the
direction of the force pointing away from the pocket. In other words, this
type of pocket repels matter. Gravity pockets can be sources of dark matter or
dark energy.

According to these ideas, there is a single origin for dark matter and dark
energy, namely, quantum tunneling of the inflaton. That there is a basically
common origin for all three sources of gravity which makes it reasonable to
expect that their quantity should be of the same order.

\section{Concluding remarks.}

We have seen that perturbative inflatons are produced from the energy of the
inflaton potential $V(\phi)$ as small loses $\Delta V$ by means of the quantum
tunneling process. These perturbative inflatons in turn disintegrate into
particles, thus contributing to the reheating process. The localized quantum
tunneling process produces an imprint on the components of the metric and
creates pockets of gravity that gravitate. The pockets can produce common
gravitation, and so be a component of dark matter, or produce repulsive
gravitation, and so be a component of dark energy. This is possible because
the Ricci tensor has terms that gravitate with either sign.

One of the immediate consequences of having a common origin for matter, dark
matter and dark energy is that their quantities should all be of the same
order, as is observed.

It was shown that the gravitational effect of a gravity pocket does not weaken
with the expansion of the Universe. As the pocket increases in size, its
strength per volume diminishes, but if we integrate over its also growing
volume, the total strength remains constant.

To understand what is happening, it is important to notice that the
second-order partials $\partial^{2}g_{\ast\ast}$and the partials square
$(\partial g_{\ast\ast})^{2}$ of the metric components are weakened by the
expansion at the same rate: \textit{both }decrease due to the common factor
$1/a^{2}$:%
\[
\frac{1}{a^{2}}\left\vert g_{\ast\ast}^{7}\frac{\partial^{2}g_{\ast\ast}%
}{\partial x^{i}\partial x^{j}}\right\vert \sim\frac{1}{a^{2}}\left\vert
g_{\ast\ast}^{6}\frac{\partial g_{\ast\ast}}{\partial x^{i}}\frac{\partial
g_{\ast\ast}}{\partial x^{j}}\right\vert .
\]
The convolutions of the metric components are not changed by the expansion,
they simply occur at a much larger scale. That is why the integral over the
whole gravity pocket gives the same amount of gravitation independently of the
expansion of the Universe.

It was shown that the expansion of the Universe does not weaken the
gravitational effect of the pockets. However, it is likely that the pockets do
weaken due to some kind of emission, perhaps of gravity quanta (whatever they
may be), but it is difficult to establish the emission rate.

Certainly there are many different kinds of pockets. Some of them may tend to
remain very small, while others will expand much faster than the expansion of
the Universe, following their own internal dynamics. These aspects are quite
complicated and hard to work out. What seems most likely is that each pocket
will evolve differently, depending mainly on its initial imprint during the
quantum tunneling process.

\bigskip

\bigskip

\begin{thebibliography}{99}                                                                                               %


\bibitem {DM}Clowe, D. \textit{et al.,} \textquotedblleft A direct empirical
proof of the existence of dark matter\textquotedblright, \textit{Astrophys.
J}. \textbf{648,} L109-L113 (2006).
\href{http://arxiv.org/abs/astro-ph/0608407}{\emph{Preprint } arXiv:0608407
[astro-ph]}

\bibitem {scaffolding}Massey, R. \textit{et al.,} \textquotedblleft Dark
matter reveals cosmic scaffolding\textquotedblright, \textit{Nature}
\textbf{445,} 286-290 (2007).
\href{http://arxiv.org/abs/astro-ph/0701594}{\emph{Preprint } arXiv:0701594
[astro-ph]}

\bibitem {MC}A primitive version of some of these ideas was presented in M.
Chaves, \textquotedblleft Dark matter from primordial metric fields and the
term (grad g\_\{00\})\symbol{94}2\textquotedblright,
\href{http://arxiv.org/abs/1107.0421}{\emph{Preprint } arXiv:1107.0421
[astro-ph]}.

\bibitem {Pokorski}S. Pokorski, \textit{Gauge field theories, 2nd. ed.},
(Cambridge, 2000).

\bibitem {H1}S. Chatrchyan et al., \textquotedblleft Observation of a new
boson at a mass of 125 GeV with the CMS experiment at the
LHC\textquotedblright, Phys. Lett. \textbf{B716}, 30-61 (2012).

\bibitem {H2}G. Aad et al., \textquotedblleft Observation of a new particle in
the search for the Standard Model Higgs boson with the ATLAS detector at the
LHC\textquotedblright, Phys. Lett. \textbf{B716}, 1-29 (2012).

\bibitem {DL}A. D. Dolgov and A. D. Linde, \textquotedblleft Baryon Asymmetry
In Inflationary Universe,\textquotedblright\ Phys. Lett. \textbf{B116}, 329 (1982).

\bibitem {AFW}L. F. Abbott, E. Farhi and M. B. Wise, \textquotedblleft
Particle Production In The New Inflationary Cosmology,\textquotedblright%
\ Phys. Lett. \textbf{B117}, 29 (1982).

\bibitem {ASTW}Albrecht AJ, Steinhardt PJ, Turner MS, Wilczek F,
\textquotedblleft Reheating An Inflationary Universe,\textquotedblright\ Phys.
Rev. Lett. \textbf{48,} 1437 (1982).

\bibitem {TB}J. H. Traschen, R. H. Brandenberger. \textquotedblleft Particle
Production During Out-Of-Equilibrium Phase Transitions,\textquotedblright%
\ Phys. Rev. \textbf{D42}, 2491 (1990).

\bibitem {DK}A. D. Dolgov, D. P. Kirilova, \textquotedblleft Production of
particles by a variable scalar field,\textquotedblright\ Sov. J. Nucl. Phys.
\textbf{51}, 172 (1990) [Yad. Fiz. \textbf{51}, 273 (1990)].

\bibitem {KLS}L. Kofman, A. D. Linde and A. A. Starobinsky, \textquotedblleft
Reheating after inflation,\textquotedblright\ Phys. Rev. Lett. \textbf{73},
3195 (1994) \& \textquotedblleft Towards the theory of reheating after
inflation,\textquotedblright\ Phys. Rev. \textbf{D56,} 3258 (1997).
\href{http://arxiv.org/abs/hep-th/9405187}{\emph{Preprint } arXiv:9405187
[hep-th]} \& \href{http://arxiv.org/abs/hep-ph/9704452}{\emph{Preprint }
arXiv:9704452 [hep-ph]}

\bibitem {GPR}B. R. Greene, T. Prokopec and T. G. Roos, \textquotedblleft
Inflaton Decay and Heavy Particle Production with Negative
Coupling\textquotedblright, Phys. Rev. \textbf{D56}, 6484 (1997).
\href{http://arxiv.org/abs/hep-ph/9705357}{\emph{Preprint } arXiv:9705357
[hep-ph]}

\bibitem {FGGKLT}G. N. Felder, J. Garcia-Bellido, P. B. Greene, L. Kofman, A.
D. Linde and I. Tkachev, \textquotedblleft Dynamics of symmetry breaking and
tachyonic preheating,\textquotedblright\ Phys. Rev. Lett. \textbf{87}, 011601
(2001). \href{http://arxiv.org/abs/hep-ph/0012142}{\emph{Preprint }
arXiv:0012142 [hep-ph]}

\bibitem {FKL}G. N. Felder, L. Kofman and A. D. Linde, \textquotedblleft
Tachyonic instability and dynamics of spontaneous symmetry
breaking,\textquotedblright\ Phys. Rev. \textbf{D64}, 123517 (2001).
\href{http://arxiv.org/abs/hep-th/0106179}{\emph{Preprint } arXiv:0106179
[hep-th]}

\bibitem {LL}A. R. Liddle and D. H. Lyth, \textquotedblleft COBE,
Gravitational Waves, Inflation and Extended Inflation\textquotedblright, Phys.
Lett. B291, 391. \href{http://arxiv.org/abs/astro-ph/9208007}{\emph{Preprint }
arXiv:9208007 [astro-th]}

\bibitem {Ma}M. Maggiore, "Gravitational wave experiments and Early Universe
Cosmology", Phys. Rep. 331, 283 (2002). See equation (160).
\end{thebibliography}
\end{document}